\documentclass[bibyear]{aa} 
\usepackage{graphicx,rotate}
\usepackage{txfonts}
\newcommand{\vm}{V_{\rm max}}
\newcommand{\vr}{V_{\rm rot}}
\newcommand{\rd}{r_{\rm d}}
\newcommand{\oii}{[O\,{\small II}]}
\newcommand{\oiii}{[O\,{\small III}]}

\begin{document} 

\title{Disk galaxy scaling relations at intermediate redshifts \\
I.~The Tully-Fisher and velocity-size relations
\thanks{Based on observations with the European Southern Observatory
Very Large Telescope (ESO-VLT), observing run IDs 65.O-0049, 66.A-
0547, 68.A-0013, 69.B-0278B, 70.B-0251A and 081.B-0107A.}
}

\author{
Asmus B\"ohm\inst{1}
\and
Bodo L.~Ziegler\inst{2}
}

\institute{Institute for Astro- and Particle Physics, Technikerstrasse 25/8, 
6020 Innsbruck, Austria.
\email{asmus.boehm@uibk.ac.at}
\and
Institute of Astronomy, T\"urkenschanzstrasse 17, 1180 Vienna, Austria.
}

\date{Received ; accepted}

 
  \abstract
   {}
{
Galaxy scaling relations such as the Tully-Fisher relation
(between maximum rotation velocity $\vm$ and luminosity)
and the velocity-size relation (between $\vm$ and disk scale length)
are powerful tools to quantify the evolution of disk galaxies with cosmic
time.
}
{
We took spatially resolved slit spectra of 261 field disk galaxies at 
redshifts up to $z \approx 1$ using the FORS instruments of the ESO
Very Large Telescope. The targets were selected from the FORS Deep Field
and William Herschel Deep Field.
Our spectroscopy was complemented with HST/ACS
imaging in the F814W filter.
We analyzed the ionized gas kinematics by extracting rotation curves
from the 2-D spectra.
Taking into account all geometrical, observational and instrumental effects,
these rotation curves were used to 
derive the intrinsic $\vm$. 
}
{
Neglecting galaxies with disturbed kinematics or insufficient spatial rotation curve extent, $\vm$ could be robustly determined for 124 galaxies covering redshifts $0.05<z<0.97$. This is one of the largest kinematic samples of distant disk galaxies to date. We compared this data set to the local $B$-band Tully-Fisher relation and the local velocity-size relation. The scatter in both scaling relations  is a factor of $\sim$\,2 larger at $z \approx 0.5$ than at $z \approx 0$. The deviations of individual distant galaxies from the local Tully-Fisher relation are systematic in the sense that the galaxies are increasingly overluminous towards higher redshifts, corresponding to an over-luminosity 
$\Delta M_B = -(1.2 \pm 0.5)$\,mag at $z=1$. This luminosity evolution at given $\vm$ is probably driven by younger stellar populations of distant galaxies with respect to their local counterparts, potentially combined with modest changes in dark matter mass fractions. The analysis of the velocity-size relation reveals that disk galaxies of a given $\vm$ have grown in size by a factor of $\sim$\,1.5 over the past $\sim$\,8\,Gyr, likely via accretion of cold gas and/or small satellites. Scrutinizing the combined evolution in luminosity and size, we find that the galaxies which show the strongest evolution towards smaller sizes at $z \approx 1$ are not those which feature the strongest evolution in luminosity, and vice versa.
}
{}


\titlerunning{Disk galaxy scaling relations}
\authorrunning{A. B\"ohm \& B.~L.~Ziegler}

\maketitle

\section{\label{intro}Introduction}

Observational studies of galaxy evolution have made great progress in 
the past decade. In the course of some projects, redshifts and
spectral energy distributions of several 10$^4$ galaxies at significant
cosmological look-back times were gained~---
two examples are the VVDS (Le F\'evre et al.~\cite{lefe05}) and
zCOSMOS (Lilly et al.~\cite{lill07}) surveys.
Other projects focused on smaller samples to conduct detailed studies of 
scaling relations which link fundamental structural and kinematical 
parameters of galaxies.
In the case of spirals, the most famous scaling relation is the   
Tully-Fisher Relation (TFR, Tully \& Fisher \cite{tull77}) which
relates the luminosity to the maximum
rotation velocity $\vm$. Equivalents of the classical, i.e.~optical 
TFR were later established,
showing that also the stellar mass $M_\ast$ 
(e.g.~Bell \& de Jong~\cite{bell01}) or 
total baryonic mass
(i.e.~stars and gas, e.g.~McGaugh et al.~\cite{mcga00}) 
correlate with $\vm$. In that sense, the optical TFR is
only a variant of a more fundamental relation between the baryonic and the dark matter content of disk galaxies.
Since dark matter dominates the total mass
budget, $\vm$ can be used to estimate the dark matter halo mass (e.g.~Mo, 
Mao \& White~\cite{mmw98b}). 
Besides luminosity, stellar or baryonic mass, also 
the disk scale length is correlated with $\vm$ 
(e.g. Mao, Mo \& White~\cite{mmw98a}), 
this is referred to as the rotation velocity - size relation (VSR).
The three parameters $\vm$, size and luminosity span a parameter space
in which disks populate a two-dimensional plane with small intrinsic scatter. 
E.g., Burstein et al.~(\cite{burs97}) utilized this parameter space
characterizing size via the effective radius;
Koda et al.~(\cite{koda00}) carried out a similar investigation using the disk
isophotal radius at $\mu_I=23.5$\,mag.
The distribution of late-type galaxies within a plane 
in this parameter space is
similar to the fundamental plane valid for early-type galaxies 
(e.g.~Dressler et~al.~\cite{dres87}).

The first observational attempts to construct the optical TFR of distant spirals
 and to quantify their evolution in luminosity 
were made almost two decades ago, e.g.~by 
Vogt et~al.~(\cite{vogt96}) and Rix et al.~(\cite{rix97}).
For many of the following years, there have been discrepant results 
from different studies 
on a possible evolution of the Tully-Fisher relation 
with cosmic time in zero point or slope.
Regarding the $B$-band TFR, 
Vogt~(\cite{vogt01})
did not find any evolution up to redshifts $z \approx 1$, while
e.g.~ B\"ohm et al.~(\cite{boeh04}),
Bamford et al.~(\cite{bamf06}) or Fern\`andez Lorenzo~(\cite{fern10})
found that disk galaxies at $z \approx 1$
were brighter by $\Delta M_B~\approx~-1$\,mag for a 
given maximum rotation velocity.
B\"ohm et al.~also discussed a possible slope change with cosmic time, in
the sense that the luminosity evolution of low-mass spirals
was stronger than that of high-mass ones.
Weiner at al.~(\cite{wein06}), on the other hand, found the opposite
evolution, i.e.~a stronger brightening in high-mass disk galaxies.
B\"ohm \& Ziegler~(\cite{boeh07}) showed that a strong evolution of the
TFR scatter could mimic an evolution in TFR slope due to 
selection effects. Kassin et al.~(\cite{kass07}) established that even 
galaxies with kinematic disturbances~--- which usually do not follow the
classical TFR~--- obey a remarkably tight correlation when
non-ordered motions are taken into account. To this end, these authors
introduced the parameter $S_{\rm 0.5}$ which combines $\vm$ and gas velocity 
dispersion, finding a constant slope over the epoch $0.1 < z < 1.2$.

Only more recently, there has been growing consent that the local TFR slope
holds at least up to redshifts about unity, for all variants of the
TFR. Some of these studies were~---
like all of the projects mentioned above~---
based on slit spectroscopy 
(e.g.~Fern\`andez Lorenzo et al.~\cite{fern10},
Miller et al.~\cite{mill11}),
others made use of Integral Field Units 
(IFUs, e.g.~Flores et al.~\cite{flor06}, 
Puech et al.~\cite{puec08}). 
Due to the time-expensive approach, samples constructed 
using IFUs usually are smaller than slit-based ones, the gain is 
a direct observability of (at least part of) the two-dimensional 
rotation velocity field.
Mismatches between photometric and kinematic center, or photometric and
kinematic position angle, can only be detected with this kind of data.

Towards high redshifts $z>1$, the usage of IFUs is a necessity, since
interaction and merger events were much more frequent at these epochs.
In effect, large fractions of high-$z$ star-forming galaxies feature
complex or disturbed kinematics.
Several surveys have been conducted with the adaptive optics-assisted 
SINFONI instrument of the ESO Very Large Telescope: 
AMAZE (Maiolino et al.~\cite{maio08}),
SINS (F\"orster-Schreiber et al.~\cite{foer09}) and
MASSIV (Epinat et al.~\cite{epin09}).
Based on data from MASSIV, e.g.~Vergani et al.~(\cite{verg12}) found only a 
small increase in stellar mass ($\sim$\,0.15\,dex on average, 
depending on the reference sample at $z=0$) between $z \approx 1.2$ 
and the present-day universe, at fixed $\vm$.
Gnerucci et al.~(\cite{gner11}) investigated 11 disks at $z \approx 3$
from the AMAZE data set, finding stellar masses smaller than locally 
by $\sim$\,1\,dex. However, 
even for this sample of regularly rotating disks, a very large
scatter was observed, and the authors concluded
that the TFR is not yet established at that cosmic epoch.
Cresci et al.~(\cite{cres09}), on the other hand, found a much smaller
evolution of 0.41 dex in log\,$M_\ast$ since $z \approx 2.2$ using SINS data.

A potential environmental dependence of the TFR has been
subject to many studies. Several authors found that field and cluster 
samples have the same TFR slope, but the scatter is increased in dense
environments
(e.g.~Moran et al.~\cite{mora07}, B\"osch et al.~\cite{boes13b}).
This probably is induced by cluster-specific interaction processes.
Tidal interactions between close galaxies can increase the star formation rate
(e.g.~Lambas et al.~\cite{lamb03}),
whereas interactions between the interstellar medium and the hot
intra-cluster medium (ram-pressure stripping), can push gas 
out of a disk galaxy and in the extreme case totally quench star formation 
(e.g.~Quilis et al.~\cite{quil00}). 
These mechanisms lead to a larger range in luminosities
at given $\vm$ and increase the fraction of perturbed gas kinematics
(e.g.~B\"osch et al.~\cite{boes13a}).
In turn, $\vm$ measurements in dense environments carry larger systematic 
errors. All these effects are likely to contribute to
the larger TFR scatter found in clusters.
However, the situation changes 
when field and cluster samples are matched in rotation
curve quality, i.e.~when galaxies are rejected that
have perturbed kinematics due to cluster-specific interactions.
It was found that
the distributions of field and cluster samples
then are very
similar in Tully-Fisher space
(e.g., Ziegler et al.~\cite{zieg03}, Nakamura et al.~\cite{naka06},
Jaff\'e et al.~\cite{jaff11}).

Concerning numerical simulations, spiral galaxies often 
had too low an angular momentum compared to observed ones
(e.g. Steinmetz \& Navarro~\cite{stei99}). 
Only more recently, it became feasible 
to simulate galaxies which over a broad 
mass range agree with the observed, local TFR,
by including
recipes for internal physics such as supernova feedback as well as external 
processes such as the ultraviolet background 
(e.g.~Governato et al.~\cite{gove07}). 
Dutton et al.~(\cite{dutt11}) used combined $N$-body simulations and
semi-analytic models to predict the evolution of several scaling
relations up to redshift $z = 4$. In these simulations, disks at $z=1$
are, at given $\vm$, brighter by $-0.9$\,mag in the $B$-band and smaller 
by $\sim$\,0.2 dex than their local counterparts. 
Based on cosmological $N$-body/hydrodynamical simulations, also
Portinari \& Sommer-Larsen~(\cite{port07}) found a $B$-band brightening by
$-0.85$\,mag at $z=1$.

Observational studies of the evolution of the VSR are
relatively scarce.
Puech et al.~(\cite{puec07}) found no change in disk sizes at given $\vm$
between $z \approx 0.6$ and $z=0$.
Vergani et al.~(\cite{verg12}) reported only a small increase of 0.12 dex
in half-light radius since $z \approx 1.2$.
Towards higher redshifts, a stronger evolution was found:
the sample presented by F\"orster-Schreiber et al.~(\cite{foer09})
yields an increase in size by a 
factor of $\sim$\,2 between $z \approx 2$ and locally 
(see Dutton et al.~\cite{dutt11}). However, this value
might be an underestimate due to the computation of disk sizes based on
H$\alpha$ half-light radii, as shown by Dutton et al.

In this paper, we will use the Tully-Fisher and
velocity-size relations to investigate the evolution of disk galaxies in
luminosity and size since redshifts $z \approx 1$.
Note that the main driver of our project is not a complete census 
of the disk galaxy population during these cosmic epochs, 
but a detailed look at only the \emph{virialized and
undisturbed} disks. Only this allows to use
scaling relations like the Tully-Fisher without the impact of kinematic
biases.
The paper is organized as follows: In Sect.~\ref{sel} we outline the selection
and observation of our sample, Sect.~\ref{redu} briefly describes the
data reduction, in Sect.~\ref{ana} we construct and analyze 
the intermediate-redshift scaling relations, Sect.~\ref{discuss} comprises the 
discussion and Sect.~\ref{concl} summarizes our main results.

In the following, we assume a flat concordance cosmology with 
$\Omega_\Lambda=0.7$, $\Omega_m=0.3$ and $H_0=70$\,km\,s$^{-1}$\,Mpc$^{-1}$.
All magnitudes are given in the Vega system.

\section{\label{sel}Sample selection and observations}

For the selection of our spectroscopic targets, we relied on two
multi-band photometric surveys: the 
FORS Deep Field (FDF; Heidt et al.~\cite{heid03}) and the William
Herschel Deep Field (WHDF; Metcalfe et al.~\cite{metc01}). 
These comprise deep imaging in the filters 
$U$, $B$, $g$, $R$, $I$, $J$, $K$ (FDF) and
$U$, $B$, $R$, $I$, $H$, $K$ (WHDF).
The filter set used in the FDF photometry is very similar to the 
Johnson-Cousins system, while the WHDF photometry is based on Harris filters;
we transformed these magnitudes to the Johnson-Cousins system via synthetic
photometry.

We applied the following criteria to construct the sample: 

\begin{enumerate}
\item total apparent brightness $R < 23$\,mag;  
\item star-forming spectral energy distribution, based on a photometric 
redshift catalog for the FDF targets (Bender et al.~\cite{bend01})
or color-color diagrams for galaxies in the WHDF, for which
no photometric redshifts were available. For the latter candidates, 
we adopted the evolutionary tracks presented by Metcalfe et 
al.~(\cite{metc01}); 
\item disk inclination angle $i > 30^\circ$ to avoid face-on disks and
ensure sufficient rotation along line-of-sight;
\item misalignment angle $\delta < 15^\circ$ between apparent major axis
and slit direction to limit geometric distortions of the observed rotation
curves. 
\end{enumerate}
Note that no selection on morphological type nor emission line strength 
was used.

The spectroscopic data were taken between September 2000 and October 2008 
with the FORS 1 \& 2 instruments of the VLT. In total, 261 disk galaxies 
were observed. All runs except the one in 2008 were carried out in multi-object 
spectroscopy (MOS) mode with straight slitlets perpendicular to the
direction of dispersion.
The observations in 2008 made use of the Mask Exchangeable Unit (MXU)
 with tilted slits, allowing to accurately place them along the
apparent major axes and achieve a misalignment angle $\delta \approx 0^\circ$.
Slit tilt angles $\theta$  were limited to 
$\lvert \theta \rvert< 45^\circ$ to ensure 
a robust sky subtraction and wavelength calibration.
We used a fixed slit width of 1.0 arcsec which resulted in a
spectral resolution of R $\approx$ 1200 (grism 600R) and a spatial scale of
0.2 arcsec/pixel for observations taken before 2002 when 
the FORS CCD was upgraded. The upgrade led to an increased sensitivity at 
wavelengths $\lambda > 7000$\,\AA, a lower readout noise and much higher
readout speed. The data taken after this upgrade feature R $\approx$ 1000 
(grism 600RI) and a scale of 0.25 arcsec/pixel.
The total integration time for all MOS setups was 2.5\,h.
Only for the MXU observations in 2008, we used a much longer integration time
of 10\,h total per target. 
This latter run was designed to particularly extend our 
sample at low luminosities, and included 
galaxies down to an apparent $R$-band
brightness of $R \approx 24$. Except for fill-up targets, all galaxies from
this run have a $B$-band absolute magnitude $M_B \gtrsim -19$.
Across all our spectroscopic campaigns, seeing conditions ranged from
0.42\,arcsec to 1.20\,arcsec FWHM, with a median of 0.76\,arcsec.

To determine structural parameters such as
disk inclination and scale length, bulge--to--disk ratios etc., we took 
HST/ACS images with the F814W filter (similar to the Cousins $I$-band).
The 6 $\times$ 6 arcmin$^2$ sky areas of the
FDF and WHDF were covered with a 2 $\times$ 2 mosaic each,
using the Wide Field Camera (0.05 arcsec/pixel) with
one orbit per pointing and
total integration times of 2360\,s (FDF) and 2250\,s (WHDF), respectively.

\section{\label{redu}Data reduction}

All reduction steps were carried out on the extracted 2-D spectra
of each exposure and each target individually. The reduction 
included bias subtraction, cosmics removal, flatfielding,
correction of spatial distortions, wavelength calibration and sky subtraction. 
Only the final, fully-reduced 2-D
spectra were co-added with a weighting factor according to
the seeing conditions during spectroscopy. 
For wavelength calibration, the dispersion relation was fitted with a 
third-order polynomial row by row; 
the median rms of these fits was 0.04\,\AA.

For the data taken with tilted slitlets in 2008, we took a different approach
and applied the improved sky subtraction method by Kelson (\cite{kels03}).
It avoids the problems arising from the transformation of a regular pixel grid
(raw spectrum) to an irregular pixel grid (rectified, wavelength-calibrated
spectrum). To this end, the night sky emission is fitted and removed after
bias subtraction and flatfielding, 
but \emph{before} the distortion correction and wavelength calibration.
Note that the difference in final data quality 
between the ``classical"  method, where sky subtraction
is carried out as the \emph{last} of all reduction steps,
and the one following Kelson is
only marginal when slitlets are oriented perpendicular to the direction of
dispersion (as is the case for our FORS data taken in MOS mode).
However, for data gained with tilted slits (MXU mode), night sky line 
residuals are strongly reduced with the Kelson approach (Kelson~\cite{kels03}).

Regarding the ACS imaging, the standard pipeline was used to carry out
bias subtraction, flatfielding
and distortion correction. We applied a filtering algorithm to finally 
combine the exposures of each pointing and remove the cosmics.

\section{\label{ana}Analysis of the scaling relations}

\begin{figure}[t]
\resizebox{\hsize}{!}{\includegraphics[angle=270]{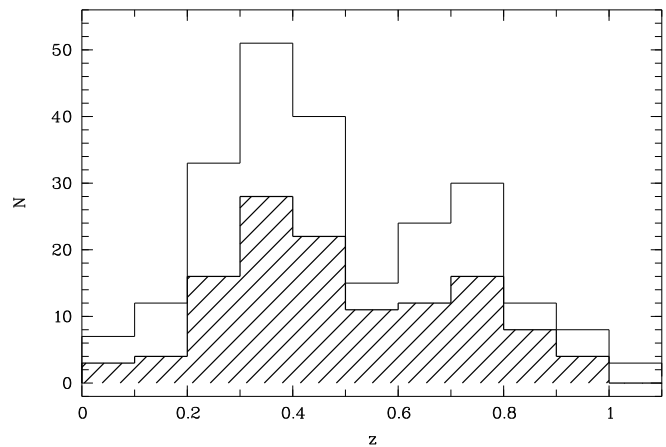}}
\caption{\label{zhis}
Redshift distribution of all galaxies in our survey (solid line, 238 galaxies; one fill-up target at $z=1.49$ is omitted in this plot) and our kinematic sample, i.e.~all galaxies used in our kinematic analysis (hashed histogram, 124 galaxies). The selection criterion for the latter sub-sample is detailed in
Sect.~\ref{anakin}.}
\end{figure}

Out of the 261 disk galaxies for which we obtained spectra, redshifts could
be determined for 238 galaxies.
These objects range from $z = 0.03$ to $z = 0.97$ (omitting an outlier
at $z=1.49$ which was a fill-up target and did not yield a $\vm$ value)
with a median of $\langle z \rangle = 0.43$.
We show the redshift distribution in Fig.~\ref{zhis}.
The gap around $z \approx 0.5$ can be 
attributed to the galaxies stemming from the FDF. 
This gap is not a ``redshift desert" due to
constraints in our spectroscopic setup; it is
also present in the distribution of the FDF
photometric redshifts, while it is \emph{absent} in the redshift distribution
of the WHDF galaxies only.
Most probably it thus is physical and a result of cosmic variance~---
both the FDF and the WHDF are deep surveys
with a relatively small field--of--view of $\approx$\,40\,arcmin$^2$ each.

\subsection{\label{analum}Absolute magnitudes}

We opt to construct the $B$-band Tully-Fisher relation in this study to
be sensitive for recent or ongoing star formation. A big advantage for this is
the multi-band imaging at hand closely matching the rest-frame $B$-band,
for any galaxy redshift in our sample.
To determine the $B$-band luminosities $M_B$, we used the
observed filter $X_{\rm obs}$ best matching the rest-frame 
$B$-band to derive the $k$-correction via synthetic photometry and 
compute the transformation $X_{\rm obs} \rightarrow B_{\rm rest}$. 
In the FDF, we used the transformation 
$B_{\rm obs} \rightarrow B_{\rm rest}$ at redshifts $z<0.25$,
$g_{\rm obs} \rightarrow B_{\rm rest}$ at $0.25<z<0.55$, 
$R_{\rm obs} \rightarrow B_{\rm rest}$ at $0.55<z<0.85$
and $I_{\rm obs} \rightarrow B_{\rm rest}$ at $z>0.85$.
For the galaxies stemming from the WHDF, we utilized
$B_{\rm obs} \rightarrow B_{\rm rest}$  at $z<0.3$, 
$R_{\rm obs} \rightarrow B_{\rm rest}$ at $0.3<z<0.7$ and
$I_{\rm obs} \rightarrow B_{\rm rest}$ at $z>0.7$.
Thanks to this approach, the $k$-correction is only weakly depending
on the spectral energy distribution of a given galaxy:
if the spectral classification would be wrong by $\Delta T = 2$ 
(which corresponds to the spectrum of, for example, an Sa galaxy mistakenly 
classified as type Sb), the
resulting systematic $k$-correction error would be $\sigma_k < 0.1$\,mag
across the whole redshift range covered by our data.

We corrected for
intrinsic absorption $A_B$ due to the dust disk following the approach by
Tully et al.~(\cite{tull98}). 
This formalism is inclination- and $\vm$-dependent:
disks which are observed more edge-on have a higher extinction than more
face-on ones, and
more massive disks, i.e.~galaxies with higher $\vm$, 
have a higher extinction than less massive ones. 

To summarize, the absolute $B$-band magnitudes were computed as
\begin{equation}
\label{MBs} 
M_B = m_X - A_X^g - DM - k_B - A_B,
\end{equation}
where
$m_X$ is the  total apparent brightness in filter $X$,
$A_X^g$ is the Galactic absorption for filter $X$, corrected using the maps
by Schlegel et al.~(\cite{schl98}),
$DM$ is the distance modulus,
$k_B$ is the $k$-correction and
$A_B$ is the correction for intrinsic dust absorption in rest-frame $B$.

\subsection{\label{anastruc}Structural parameters}

We derived the structural parameters of the galaxies~---
disk inclination, disk scale length $\rd$, bulge--to--total ratio $B/T$ 
etc.~--- on the HST/ACS images 
using the GALFIT package by Peng et al.~(\cite{peng02}). 
It allows to fit multiple two-dimensional surface brightness profiles
simultaneously to the galaxy under
scrutiny as well as to neighboring objects
(the importance of such simultaneous fits is discussed, 
e.g., in H\"au{\ss}ler et al.~\cite{haeu07}).
GALFIT requires an input Point Spread Function (PSF) for the convolution
of the model profiles. 
We constructed PSFs for the FDF and WHDF separately 
from $\sim$\,20 unsaturated stars in each field. Both have a FWHM
of 0.12\,arcsec, corresponding to a spatial resolution of
$\sim$\,0.7\,kpc at $z=0.5$.

The surface brightness profiles of all galaxies in our sample were fitted 
with two different setups:
i) a single S\'ersic profile with free index $n_{\rm ser}$ or
ii) a two-component model with an exponential profile for the disk and
a S\'ersic profile with fixed index $n_{\rm ser}=4$ for the bulge.
The best-fit parameters from method i) were used as initial guess values 
for method ii).

All fit residuals were visually inspected, and in a few cases, constraints
on parameters like e.g.~bulge effective radius were necessary to avoid a local
$\chi^2$ minimum in the fitting process.
For our analysis, we mostly used the disk parameters from the bulge/disk 
decomposition, i.e.~method ii). Only in a few cases where the two-component
fit of an evidently bulgeless disk was not converging, we kept the
parameters from the single S\'ersic fit. 
We stress that for the analysis presented 
here, the most important parameters are the inclination $i$, 
position angle $\theta$ and scale length $\rd$ of
the disk. These showed only small differences between the two fitting 
methods. This is mainly because the vast majority of the FDF/WHDF disks have
only small bulges or even no detectable bulge at all; the median 
bulge--to--total ratio is $\langle B/T \rangle = 0.06$.

GALFIT only returns random errors on the best-fit parameters. These are
very small ($<$1\,\%) throughout our sample. To gain a more realistic estimate of the systematic errors on $\rd$, we can rely on our own previous analysis of HST/ACS images using GALFIT in B\"ohm et al.~(\cite{boeh13}). In that work, we investigated the impact of an Active Galactic Nucleus on the morphologies of host galaxies at redshifts $0.5 < z < 1.1$ as quantified with GALFIT. For a \emph{negligible} central point source, we found a typical systematic error of 20\,\% on galaxy sizes (see Fig.~7d in B\"ohm et al.~\cite{boeh13}). This value hence represents the systematic size error for galaxies with the light profiles of pure disks or disks with only weak bulges; this is the case for the vast majority of galaxies in our sample. We therefore adopt this error on $\rd$ in the following.

It is well-known that the observed disk scale length depends on the wavelength 
regime (see, e.g., de Jong~\cite{dejo96}), in the sense that $\rd$ is smaller in
redder filters. The effect is rather small in the F814W filter for the 
redshifts covered by our sample and at maximum
corresponds to an overestimate of $\rd$ by 11\% for the highest-redshift galaxy
in our sample, at $z=0.97$. 
We corrected (i.e.~reduced) all measured disk scale lengths for this effect~---
depending on a given galaxy's redshift~--- to make them directly comparable.\\

\subsection{\label{anakin}Kinematics}

Rotation curves (rotation velocity as a function of radius)
were extracted from the two-dimensional spectra by fitting Gaussian profiles
to the emission lines stepwise along the spatial axis. We used a boxcar of three
pixels, averaging over a given spectral row and its two adjacent rows. 
This approach increases the $S/N$ without a loss in spatial resolution~--- 
the boxcar size corresponds to 0.6\,arcsec in the FDF spectra and 0.75\,arcsec 
in the WHDF spectra; both values are below the average seeing during 
spectroscopy. All detected emission lines were used, and the rotation curve 
with the best $S/N$ and largest spatial extent was used as reference in the 
further analysis. The typical error on the rotation velocity at a given 
galactocentric radius is 10-20\,km/s.
Approx.~half of the reference rotation curves stem from the \oii\ doublet, the 
other half is based on either the \oiii, H$\beta$ or H$\alpha$ line.
The vast majority of the rotation curves extracted from different emission lines
agreed within the errors of the Gaussian fits, in the sense that the 
rotation velocities at given radius agreed within their errors.

The derivation of the maximum rotation velocity $\vm$ of distant galaxies is a 
challenging task. At $z \approx 0.5$, 
the half-light radius of a Milky Way-type galaxy
is similar to the slit width and the typical seeing FWHM. This leads to strong
blurring effects in the observed rotation curves. 
It is mandatory to take these effects into account to avoid underestimates of 
$\vm$. To tackle this problem, we introduced a methodology
that simulates all steps of the observation process, from the
intrinsic 2-D rotation velocity field to the extracted 1-D rotation curve.

The \emph{intrinsic rotation velocity field}~--- unaffected by any geometrical,
atmospherical or instrumental effects~--- is generated assuming
a linear rise of the rotation velocity $\vr(r)$ at radii $r<r_t$, where
 $r_t$ is the co-called turnover radius, 
and a convergence of $\vr(r)$ to a constant value $\vm$
at $r>r_t$ 
(Courteau~\cite{cour97}). 
We adopt a turnover radius equal to the scale length 
$r_{d,gas}$ of the emitting ionized gas disk; $r_{d,gas}$, in turn, 
is computed from the stellar
disk scale length $\rd$ following Ryder \& Dopita (\cite{ryde94}). 
The intrinsic velocity field is then transformed into a 
\emph{simulated rotation curve} including 
the following effects: 
\begin{enumerate}
\item disk inclination angle $i$; 
\item misalignment angle $\delta$ between the slit orientation
and the apparent major axis;
\item seeing during spectroscopy;
\item luminosity profile weighting perpendicular to the slit direction;
\item blurring effect due to the slit width in direction of
dispersion~--- the optical equivalent to ``beam smearing" in radio
observations. 
\end{enumerate}
The simulated rotation curve is then fitted to the observed
one to infer the intrinsic maximum rotation velocity $\vm$. 

In this paper, we use $\vm$
as the only free parameter in the rotation curve fitting process.
The other parameters are held fixed, based on the observed disk inclination,
position angle, scale length, etc.
For testing purposes, 
we utilized $r_t$ as a second fitting parameter and found that
the results on $\vm$ agreed with the single-parameter fits 
within the errors for $\sim$\,90\% of the objects.
This is probably due to the strong blurring effects which limit or even 
erase information on the rotation curve shape at small galactocentric radii.

We also investigated whether our results are depending on the intrinsic 
topology of $\vr(r)$. 
When we, instead of the Courteau et al.~(\cite{cour97}) parametrization,
use the more complex Universal Rotation Curve (URC) shape introduced by 
Persic et al.~(\cite{pers96}), the recomputed $\vm$ values agree with the
Courteau-based ones for 96\% of our sample.
In brief, the URC was inferred from $>$\,1000
observed rotation curves of local spiral galaxies; 
it comprises a mass-dependent velocity gradient.
Very low-mass spirals show an increasing rotation velocity even at large radii,
whereas the rotation curves of very high-mass spirals are moderately 
declining in that 
regime. The fact that our results on $\vm$ are not sensitive to the choice of 
the intrinsic rotation curve shape probably has two reasons.
Firstly, our sample mainly covers intermediate masses, where the URC does not
introduce a velocity gradient at large galactocentric radii.
Secondly, the radial extent of the
observed rotation curves~--- typically four disk scale lengths~---
might be too small to robustly detect a potential velocity 
gradient in the outer disk.

To ensure a robust analysis of the scaling relations, we visually inspected 
all rotation curves to identify those which
a) show kinematic perturbations or 
b) have an insufficient spatial extent and do not probe
the regime where $\vr(r)$ converges to $\vm$; in extreme cases these 
curves show solid-body rotation.
During this visual inspection, we rejected 101 objects from the data set of
238 galaxies with determined redshifts, corresponding to a fraction of
42\,\%. 
Another 13 galaxies were not considered for further analysis because the rotation curve fitting yielded a very large error on $\vm$, with a relative error $\sigma_{\rm vrel} = \sigma_{\rm vmax} / \vm > 0.5$ (errors on $\vm$ stem from fitting the synthetic rotation curves to the observed rotation curves via $\chi^2$ minimization). The excluded 13 galaxies are relatively faint (median total apparent $R$-band brightness $\langle R \rangle = 22.42$, compared to $\langle R \rangle = 21.81$ for the rest of the sample with derived $\vm$) and their rotation curves show slight asymmetries; otherwise these galaxies have properties similar to the remaining sample.

The 124 galaxies with a robust $\vm$ constitute our kinematic sample.
It covers a range $25\,{\rm km/s} < \vm < 450\,{\rm km/s}$ with a median value
of $\langle \vm \rangle = 145$\,km/s.
The median of the relative error of the maximum rotation velocities is 
$\langle \sigma_{\rm vrel} \rangle = 0.19$. 
This error in general is larger towards lower
values of $\vm$, 
i.e.~towards low-mass galaxies, and smaller towards higher
values of $\vm$, i.e.~towards high-mass galaxies.
Table~\ref{datatab} lists the main parameters of ten galaxies as an example (the full table comprising 124 galaxies is  electronically available):
object ID, redshift, maximum rotation velocity, $B$-band absolute magnitude, and disk scale length.

\begin{table}
\caption{Main galaxy parameters}
\label{datatab}
\begin{tabular}{ccccc}
\hline\hline 
ID & $z$ & $\vm$ & $M_B$\tablefootmark{a} & $\rd$\tablefootmark{b} \\
 & & [km/s] & [mag] & [kpc] \\ 
\hline 
698 & 0.5663 & 155 $\pm$  46 & -20.43 $\pm$ 0.09 &  2.90 $\pm$ 0.58 \\
745 & 0.6986 & 290 $\pm$  34 & -21.77 $\pm$ 0.07 &  2.83 $\pm$ 0.57 \\
759 & 0.0718 &  \hspace{5pt}65 $\pm$  15 & -17.14 $\pm$ 0.09 &  1.06 $\pm$ 0.21 \\
762 & 0.4343 & 280 $\pm$  84 & -20.70 $\pm$ 0.09 &  1.63 $\pm$ 0.33 \\
814 & 0.6491 & 145 $\pm$  55 & -21.09 $\pm$ 0.09 &  4.10 $\pm$ 0.83 \\
832 & 0.5477 & \hspace{5pt}260 $\pm$ 104 & -19.55 $\pm$ 0.09 &  2.29 $\pm$ 0.46 \\
868 & 0.4573 & 135 $\pm$  18 & -20.71 $\pm$ 0.09 &  3.18 $\pm$ 0.64 \\
870 & 0.2775 &  90 $\pm$   4 & -20.23 $\pm$ 0.07 &  2.50 $\pm$ 0.50 \\
876 & 0.8324 & 165 $\pm$  18 & -21.54 $\pm$ 0.09 &  4.68 $\pm$ 0.95 \\
878 & 0.2128 & 140 $\pm$  50 & -19.09 $\pm$ 0.09 &  3.65 $\pm$ 0.73 \\
\hline 
\end{tabular}
\tablefoot{
Ten objects shown as examples. The full table with 124 galaxies is electronically available.
\tablefoottext{a}{Rest-frame $B$-band absolute magnitude, corrected for intrinsic absorption following Tully et al.~(\cite{tull98}).}
\tablefoottext{b}{Disk scale length from F814W HST/ACS imaging, corrected for rest-frame wavelength dependence following de Jong~(\cite{dejo96}).}
}
\end{table}

\subsection{\label{anatfr}The Tully-Fisher relation}

\begin{figure*}[t]
\centering
\includegraphics[angle=270,width=14cm]{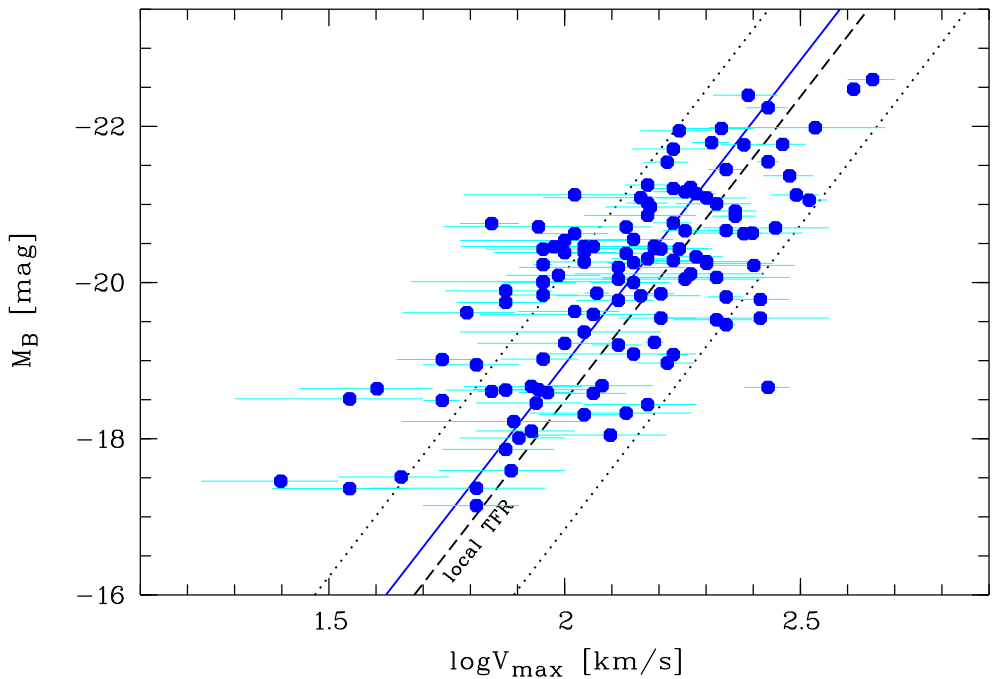}
\caption{\label{tf1}
Rest-frame $B$-band Tully-Fisher diagram, showing a 
comparison between our sample of 124 disk galaxies at a median redshift
$\langle z \rangle \approx 0.5$ (solid
symbols) and the local Tully-Fisher relation as given 
by Tully et al.~(\cite{tull98}; the
black dashed line indicates the fit to the local data 
(not shown in this figure); 
the dotted lines give the local 3\,$\sigma$ scatter). 
The blue solid line shows the fit to the distant sample
(with the slope fixed to the local value), which is offset from the local 
Tully-Fisher relation towards higher luminosities by 
$\langle \Delta M_B \rangle = -0.47 \pm 0.16$\,mag.
}
\end{figure*}

We show the distribution of our kinematic sample in rotation velocity - luminosity space in Fig.~\ref{tf1}. Our data representing a median redshift of $z=0.45$ 
are shown as solid symbols in comparison to the local Tully-Fisher relation as given by Tully et al.~(\cite{tull98}):
\begin{equation}
M_B = -7.79 \cdot \log \vm -2.91, \label{localtfr}
\end{equation}
which is shown as a dashed line in Fig.~\ref{tf1}; the dotted lines indicate the 3\,$\sigma$ scatter of the local data. Local and distant galaxies are consistently corrected for intrinsic dust absorption following the approach of Tully et al.~(\cite{tull98}). The redshift distribution of our kinematic sample of 124 galaxies is shown in Fig.~\ref{zhis}. It is very similar to the full sample of 238 galaxies with determined redshifts. The majority of the distant galaxies are located within the 3\,$\sigma$ limits of the local TFR (see Fig.~\ref{tf1}). Interestingly, almost all galaxies with rotation velocities $\log \vm<2$ fall on the high-luminosity side of the local relation, and the eight galaxies with $\log \vm<1.8$ are even located above the 3\,$\sigma$ limits of the local TFR. Three possible explanations have to be considered.

Firstly, $\vm$ 
could be underestimated for low-mass disk galaxies: optical 
rotation curves of low-mass disks in the local universe often
have a positive rotation velocity gradient even at the largest 
covered radii (e.g., Persic et al.~\cite{pers96}). 
As outlined in Sect.~\ref{anakin}, we
did not find significant changes in $\vm$ when adopting
intrinsic rotation velocity fields with positive gradients in low-mass
galaxies (using the prescriptions presented in Persic et al.).
However, the spatial extent of the rotation curves in our sample
(as well as other samples at similar redshifts)
is typically three to four times the optical disk scale length, which
might be insufficient to constrain potential $\vr$ gradients.

Secondly, the observations might hint to a mass-dependent evolution
in luminosity that is larger for low-mass disk galaxies than for
high-mass ones. This could be interpreted as a manifestation of 
the down-sizing phenomenon (e.g., Kodama et al.~\cite{koda04}).
A third interpretation would be that the apparent mass-dependency is
only \emph{mimicked} by the impact of the magnitude limit of our sample.
It has been demonstrated by numerous authors (e.g.~Willick~\cite{will94})
that, towards lower $\vm$, 
mainly galaxies on the high-luminosity side
of a parent unbiased Tully-Fisher distribution enter an observed sample.
For a detailed discussion of this magnitude limit effect, please see
Appendix~\ref{slope}. 

While the first interpretation outlined above is a purely kinematical one, the other two only concern the luminosity, either in the rank of a physical effect (down-sizing) or a selection effect (magnitude bias). Because our targets have been selected for observation using a limit in apparent magnitude~--- which is a common and well-motivated approach~--- the third scenario seems much more likely than the second one. We again refer the reader to Appendix~\ref{slope} for a further discussion of this topic.
Towards higher $\vm$, however, most rotation curves should be flat even beyond the radii probed by our data, provided that the internal mass distribution in intermediate-$z$ and local disk galaxies is similar (e.g.~Sofue \& Rubin~\cite{sofu01}). Furthermore, the impact of the magnitude limit becomes negligible towards higher $\vm$ (e.g.~Giovanelli et al.~\cite{giov97}).

Note that overestimated luminosities at low $\vm$ 
are highly unlikely,
as the ground-based photometry is very deep and we  
compute $M_B$ from the filter which best matches the 
$B$-band in
rest-frame, assuring very small total errors on the absolute magnitudes
(combined random and systematic errors are 
$\sigma_{\rm MB}<0.12$\,mag for all galaxies).
The offsets of the slow rotators from the local TFR can also not be
attributed to an evolution in redshift: the median redshift of
these galaxies ($\langle z \rangle = 0.37$ for the eight
TFR outliers at $\log \vm<1.8$)
is lower than that of the remaining sample
($\langle z \rangle = 0.45$). 
We will revisit the question of any redshift dependencies further below.

Computing the offsets of individual galaxies
from the local TFR (as given in Eq.~\ref{localtfr}) via
\begin{equation}
\Delta M_B = M_B + 7.79 \cdot \log \vm + 2.91, \label{tfoff}
\end{equation}
we find that the distant galaxies are,  for a given $\vm$, more
luminous than their local counterparts, with a median value
$\langle \Delta M_B \rangle = -0.47 \pm 0.16$\,mag 
(shown as a solid line in Fig.~\ref{tf1}).
The scatter in $\Delta M_B$, i.e., the scatter of the distant TFR under the
assumption of the local slope, is  
$\sigma_{\rm obs}=1.28$\,mag,
which is 2.3\,$\times$\,$\sigma_{\rm obs}$ of the local $B$-band TFR,
for which Tully et al.~give 
$\sigma_{\rm obs}=0.55$\,mag.
The fixed-slope fit to the distant data shows a significant
over-luminosity of the distant disk galaxies. However, the
scatter of the distant sample is large, more than twice the local value,
and the vast majority of the distant galaxies are located within the
3\,$\sigma$ limits of the local TFR.
This large scatter 
might be in part due to the broad range in redshifts
(see Fig.~\ref{zhis}), as the evolution in luminosity might
depend on look-back time (other possible sources of increased 
scatter will be discussed in Sect.~\ref{discuss}).
We will therefore now shift
our focus from the \emph{global} evolution in luminosity to the evolution of
\emph{individual} galaxies.

\begin{figure*}[t]
\centering
\includegraphics[angle=270,width=14cm]{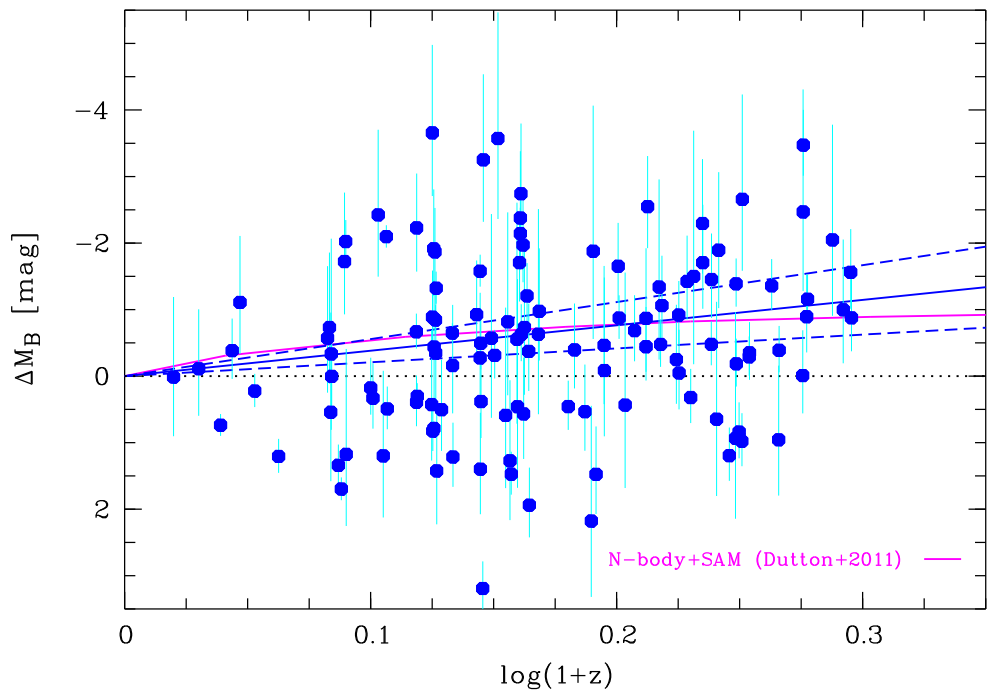}
\caption{\label{tf2}
Offsets $\Delta M_B$ of the galaxies in our sample from the local Tully-Fisher
relation, displayed as a function of redshift. 
The galaxies show increasing over-luminosities towards longer look-back times,
reaching $\Delta M_B = -(1.2\pm0.5)$\,mag 
at $z=1$ according to a linear fit, depicted
as a blue solid line and dashed lines indicating the 1\,$\sigma$ error range.
The observed luminosity evolution is in good agreement with 
predictions from numerical simulations by
Dutton et al.~(\cite{dutt11}; solid magenta line) who find 
$\Delta M_B = -0.9$ at redshift unity.
The dotted line illustrates no evolution in luminosity.
}
\end{figure*}

We show the offsets from the local TFR, computed following Eq.~\ref{tfoff},
as a function of redshift in Fig.~\ref{tf2}.
The errors on $\Delta M_B$ are computed via error propagation from the
errors on $\vm$ and $M_B$.
The TFR offsets $\Delta M_B$ show an 
evolution towards higher luminosities at higher redshifts.
This is confirmed by a linear fit to the full sample, which yields
\begin{equation}
\Delta M_B = -(3.82 \pm 1.74) \cdot \log (1+z) - (0.15 \pm 0.32). \label{tfoff1}
\end{equation}
At given $\vm$, disk galaxies at $z=1$ are on average more luminous by 
$\Delta M_B = -1.2\pm0.5$\,mag according to this fit.
As shown in Fig.~\ref{tf2}, the fit to our data is in 
good agreement with Dutton et al.~(\cite{dutt11}) who used
combined $N$-body simulations and semi-analytical models. 
With $N$-body/hydrodynamical simulations, 
Portinari \& Sommer-Larsen~(\cite{port07}) find  
$\Delta M_B~=~-0.85$\,mag at $z=1$; in excellent agreement
with the Dutton et al.~value of $\Delta M_B = - 0.9$\,mag, and our own result.
The non-linear evolution of the model prediction (see Fig.~\ref{tf2})
can, however, not be confirmed with our
data. A second-order polynomial fit agrees with the linear one to within
$\pm 0.1$ \,mag throughout the probed redshift range. \\

\subsection{\label{anavsr}The velocity-size relation}

We will now investigate the rotation velocity-size relation. 
For this purpose, we utilize the data on $\sim$\,1100 local disk galaxies
by Haynes et al.~(\cite{hayn99b}). Since the electronically available
disk sizes are given as isophotal diameters at 23.5\,mag $I$-band surface
brightness, we transformed these into disk scale lengths assuming an
average central surface brightness of $\mu_I=19.4$\,mag, 
as given in Haynes et al.~(\cite{hayn99a}).

With a bisector fit~--- a combination of two least–square
fits with the dependent and independent variables interchanged
(e.g.~Isobe~\cite{isob90})~--- to the local data, we find
\begin{equation}
\log \rd = (1.35 \pm 0.04) \cdot \log \vm - (2.41 \pm 0.08). \label{localvsr}
\end{equation}

\begin{figure*}[t]
\centering
\includegraphics[angle=270,width=14cm]{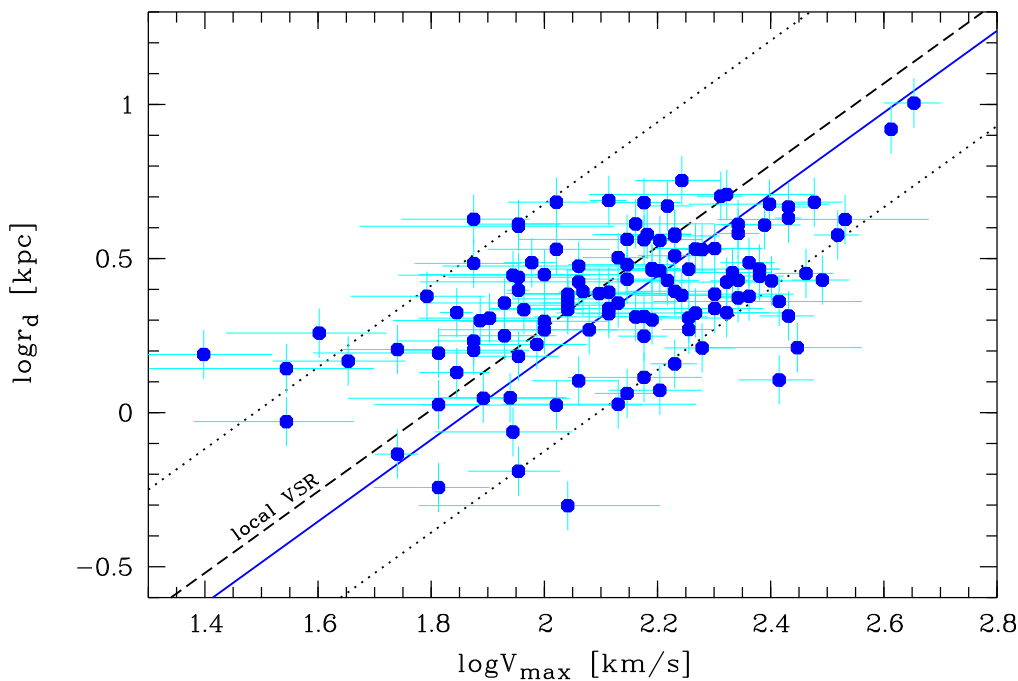}
\caption{\label{vsr1}
Velocity-size relation (VSR): disk scale length $\rd$ as a function of 
maximum rotation velocity $\vm$. We show a 
comparison between our $\langle z \rangle \approx 0.5$ data (circles)
and the local VSR as found with data from 
Haynes et al.~(\cite{hayn99b}; the black dashed line 
indicates the fit to the local data (not shown in this figure)
and the dotted lines depict the 3\,$\sigma$ scatter). 
The blue solid line shows the fit to the distant sample which is offset
to smaller disk sizes by 
$\langle \Delta \log \rd \rangle = -0.10 \pm 0.05$\,dex.
}
\end{figure*}

Fig.~\ref{vsr1} shows our data compared to the local VSR as defined by
Eq.~\ref{localvsr}. A few galaxies at the lowest $\vm$ are scattered towards large radii in VSR space. These are galaxies that also fall above the 3\,$\sigma$-limits of the local TFR. Either $\vm$ is underestimated for these galaxies, or they are more luminous \emph{and} larger than to be expected for their maximum rotation velocities.

We compute the offsets from the local VSR as
\begin{equation}
\Delta \log \rd = \log \rd - 1.35 \cdot \log \vm + 2.41,
\end{equation}
which yields a median of $\langle \Delta \log \rd \rangle = -0.10 \pm 0.05$ (displayed as a solid line in Fig.~\ref{vsr1}). The scatter in the distant sample is $\sigma_{\rm obs} = 0.27$\,dex; $2.1\,\times$ larger than the local value of $\sigma_{\rm obs}=0.13$\,dex.

\begin{figure*}
\centering
\includegraphics[angle=270,width=14cm]{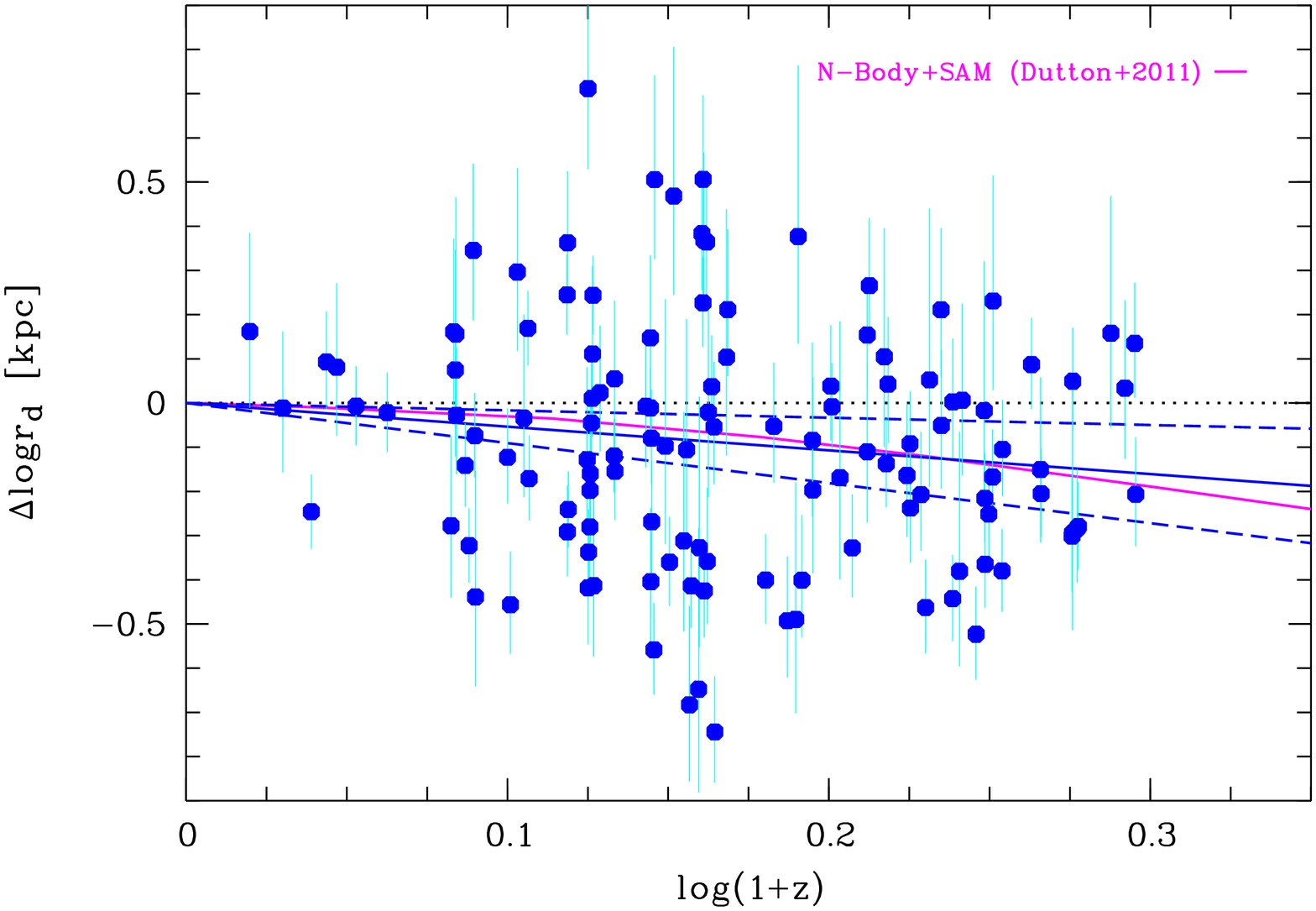}
\caption{\label{vsr2}
Offsets $\Delta \log \rd$ of our galaxy sample from the local 
velocity-size relation (see Fig.~\ref{vsr1}) as a function of redshift.
Negative values in $\Delta \log \rd$ correspond to smaller sizes at
given maximum rotation velocity $\vm$. We find successively smaller disks
towards higher redshifts. A linear fit to our data 
(displayed as a blue solid line and dashed lines indicating the 
1\,$\sigma$ error range) shows that disk galaxies
at $z=1$ were smaller than their local counterparts by a factor of $\sim$\,1.5.
For comparison, the solid magenta line shows predictions from simulations
by Dutton et al.~(\cite{dutt11}). The dotted line illustrates no evolution
in size.
}
\end{figure*}

We now again focus on a potential evolution with look-back time. Fig.~\ref{vsr2} shows the VSR offsets as a function of redshift. Errors on $\Delta \log \rd$ are propagated from errors on $\vm$ and $\rd$. Using a linear fit, we find
\begin{equation}
\Delta \log \rd = -(0.54 \pm 0.37) \cdot \log (1+z) + (0.00 \pm 0.07). \label{vsroff1}
\end{equation}
Disk galaxies at $z=1$ therefore were smaller than locally by a factor of 
$1.45^{+0.42}_{-0.33}$ on average, for a given $\vm$. This reflects ongoing disk growth with cosmic time, as expected for hierarchical structure formation (e.g., Mao, Mo \& White~\cite{mmw98a}). 
As a test, we also applied a second-order polynomial fit to the VSR offsets, not finding a significant second-order term and large errors. Similar to the TFR offsets, we therefore can not reproduce the non-linear evolution with cosmic time predicted by the Dutton et al.~(\cite{dutt11}) models. \\

\section{\label{discuss}Discussion}

To interpret the observed evolution of our disk galaxy sample in Tully-Fisher
space, we need to recall that several processes might occur between 
previous cosmic epochs and the local universe, and it could be a combination
of these processes that governs the evolution in $\Delta M_B$.
For the following discussion, we remind the reader that the maximum rotation velocity $\vm$ is a proxy for the total (virial) mass of a disk galaxy such that $M_{\rm vir}\propto\vm^3$ (e.g., van den Bosch~\cite{vdb02}). Since the virial mass is dominated by dark matter, and the optical luminosity is dominated by stellar light, the optical TFR reflects a fundamental interplay between dark and baryonic matter.

The stellar populations of distant galaxies are likely to have a younger mean age; this translates into a lower stellar $M/L$ ratio and, in turn, a lower total (baryonic and dark matter) $M/L$ ratio than locally.
At given $\vm$, distant galaxies thus would be expected to be more luminous
than local ones, particularly in the $B$-band considered here which is
sensitive to high-mass stars with short lifetimes. 
The gas mass fractions probably are higher towards higher redshifts
(e.g., Puech et al.~\cite{puec10}), corresponding to higher total 
$M/L$ ratios and TFR offsets $\Delta M_B>0$.
Furthermore, as less gas has been converted into stars, 
stellar masses $M_\ast$ might be lower at given $\vm$, and this, in turn, 
would lead to lower luminosities and, again, positive TFR offsets.
The observational census on the evolution in stellar mass is somewhat unclear
at $z<1$:
e.g., Puech et al.~(\cite{puec08}) find a decrease in stellar mass 
of $\Delta M_\ast = -0.36^{+0.21}_{-0.06}$\,dex up to $z=0.6$, while 
Miller et al.~(\cite{mill11}) give a small and statistically 
insignificant evolution of 
$\Delta M_\ast = -0.04\pm0.07$\,dex up to $z=1$. 
Using simulations, Dutton et al.~(\cite{dutt11}) 
predict a growth in stellar mass
by $\sim$\,0.15\,dex between $z=1$ and $z=0$ at fixed $\vm$.

Note that we do not imply that, for an \emph{individual} galaxy, the evolutionary path in TFR space would at all times be purely in luminosity between $z=1$ and $z=0$. E.g., if a disk galaxy would undergo a minor merger, the remnant would, after relaxation of the kinematical disturbances, most likely be located at a higher $\vm$ and a higher luminosity than before the encounter. During or shortly after the minor merger, such a galaxy would not enter our kinematic sample, even if it was covered by our observations, because it would feature a disturbed rotation curve. This is just to give an example that we do not claim disk galaxy evolution at $z<1$ to solely proceed parallel to the luminosity axis in TFR space.
A minor merger with a low-mass satellite would probably lead to an increase
of the total $M/L$ ratio, as dwarf galaxies have higher 
dark matter mass fractions than galaxies in the 
$M^\ast$ regime (e.g.~Moster et al.~\cite{most10}).
Lacking data on gas mass fractions, we can not infer dark matter mass 
fractions for our sample. Regarding the ratio 
$M_\ast/M_{\rm vir}$ between stellar mass
and virial mass, Conselice et al.~(\cite{cons05}) observationally found
a mild decrease between $z \approx 1$ and $z \approx 0$, corresponding to a slight increase in total $M/L$ ratio. 
Using semi-analytic models, Mitchel et al.~(\cite{mitc16}) inferred a basically constant $M_\ast/M_{\rm halo}$ at $0<z<1$ (cf.~their Fig.~2).

Out of the processes described above, the younger stellar populations towards higher redshifts~--- potentially in combination with slightly lower dark matter mass fractions~--- are likely dominating in our sample, since we find an increase in luminosity of $\Delta M_B=-1.2 \pm 0.5$ at given $\vm$ towards $z=1$, hence a \emph{decrease} in total $M/L$ ratio. Our result agrees with previous observational findings e.g.~by Bamford et al.~(\cite{bamf06}) or Miller et al.~(\cite{mill11}), and also with predictions from simulations by Portinari \& Sommer-Larsen~(\cite{port07}) and Dutton et al.~(\cite{dutt11}). A stellar population modeling of 108 galaxies from our sample (Ferreras et al.~\cite{ferr14}) revealed the well-known downsizing effect: the distant high-mass disk galaxies began forming their stars at higher redshifts and on shorter timescales than the low-mass ones. 

The observed evolution in disk scale length (Fig.~\ref{vsr2}) reflects
the growth of disks with ongoing cosmic time. Such an evolution is expected
in an LCDM cosmology with hierarchical structure growth.
Based on theoretical considerations, it has been predicted already 
by Mao, Mo \& White~(\cite{mmw98a}). In the cosmology adopted here, 
the computations by these authors correspond to a disk size increase 
by a factor of $\sim$\,1.9 between $z=1$ and $z=0$ at given $\vm$;
larger than what we find, but almost in agreement within the errors.
The fit to the observed evolution given in Eq.~\ref{vsroff1} corresponds
to $\Delta \log \rd = -0.16$\,dex at $z=1$. This is
a slightly stronger evolution in size than given in the 
observational study of Vergani et al.~(\cite{verg12}; these authors derived
$\Delta \log \rd = -0.12$\,dex at $z = 1.2$),
and a slightly smaller evolution than predicted with semi-analytical
models by Dutton et 
al.~(\cite{dutt11}), who find $\Delta \log \rd = -0.19$\,dex at $z = 1$.
Candidate processes to explain the disk growth towards $z=0$ are accretion 
of cold gas or minor mergers with small satellites. Any 
major mergers in the cosmic past of galaxies in our kinematical sample must have
occurred several Gyr ago so that the merger remnant could regrow a rotationally
supported disk (e.g., Governato et al.~\cite{gove09})~---
major merger remnants are kinematically cold only in special pre-merger
configurations (e.g., Springel \& Hernquist~\cite{spri05}).

Note that the Haynes et al.~(\cite{hayn99b}) sample that we used as a local
VSR reference comprises disk scale lengths
derived in the $I$-band, the response function of which is very similar to that
of the F814W filter used for our HST imaging. Since we
corrected all scale lengths to rest-frame in our data set, local reference
and distant galaxies can be directly compared.
If the required correction for the wavelength dependence of $\rd$
(following de Jong~\cite{dejo96}) would be omitted, our sample would yield 
a weaker size evolution, corresponding to 
$\Delta \log \rd = -0.11$\,dex at $z=1$ for a given $\vm$.

Our analysis so far has shown that the disk galaxy population as a whole is evolving in luminosity and size over the redshifts covered by our data. However, we do not know yet how this \emph{combined} evolution proceeds for individual galaxies. Do the galaxies with the strongest evolution in size also show the strongest evolution in luminosity, or is the picture more complex?

\begin{figure*}
\centering
\includegraphics[angle=270,width=17cm]{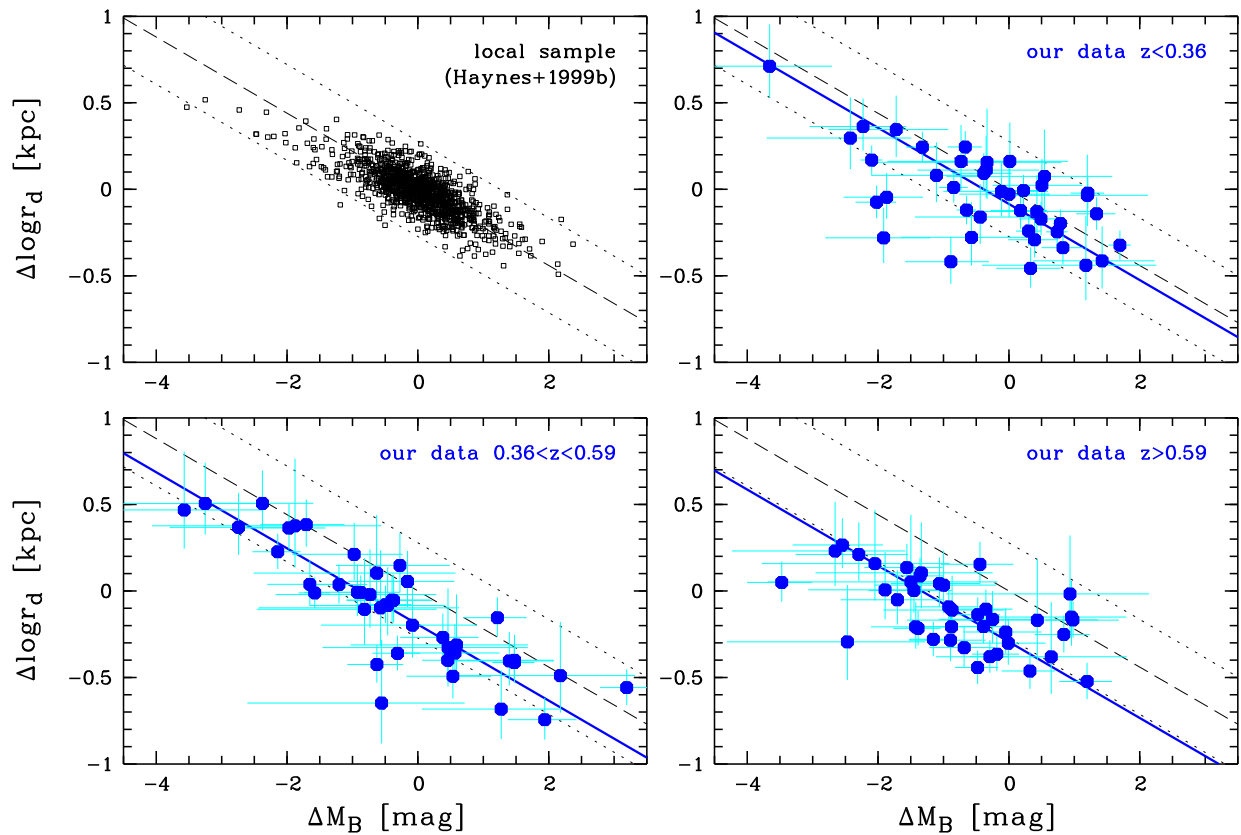}
\caption{\label{resres}
Combined look at the offsets $\Delta M_B$ from the Tully-Fisher relation and the offsets $\Delta \log \rd$ from the velocity-size relation.
\emph{Top left:} local sample from Haynes et al.~(\cite{hayn99b}; no error bars shown for clarity), showing a correlation between the two parameters (the dashed and dotted lines show the fit and 3\,$\sigma$ scatter). 
\emph{Top right, bottom left and bottom right:} distribution of our data in three redshift bins comprising 41-42 galaxies each, with a fixed-slope fit indicated by a solid line. The local fit and 3\,$\sigma$ scatter are displayed for comparison in each panel. These figures show the combined evolution in luminosity and size towards higher higher redshifts. See text for details.}
\end{figure*}

To look further into this, we compare the offsets from the Tully-Fisher relation $\Delta M_B$ to the offsets from the velocity-size relation $\Delta \log \rd$ as shown in Fig.~\ref{resres}. We will first consider the situation in the present-day universe: the upper left panel shows the local sample from Haynes et al.~(\cite{hayn99b}). The offsets $\Delta M_B$ of these galaxies are computed from the local TFR as derived with the same sample, and also the offsets $\Delta \log \rd$ from the local VSR (as given in Eq.~\ref{localvsr}) are based on the same sample. By construction, the median of both parameters is zero.
At first glance, it might be surprising that $\Delta M_B$ and $\Delta \log \rd$ are correlated: the dashed line depicts a fit to the local data; the dotted lines illustrate 3\,$\times$ the local scatter, which in terms of $\Delta \log \rd$ is $\sigma = 0.09$\,dex.
This correlation can be understood as a result of the fact that disk galaxies populate a \emph{plane} in the three-dimensional parameter space span by maximum rotation velocity, luminosity and size (e.g., Koda et al.~\cite{koda00}). Since the TFR is a projection of this fundamental plane, any deviation of a galaxy in the local universe from the TFR depends partly on its size, or, more precisely, on its position in luminosity-size space. 
In other words, \emph{Fig.~6 is equivalent to an edge-on view on the fundamental plane of disk galaxies in the direction of}~--- but not parallel to~--- \emph{the $\vm$ axis.}

We divide our sample into three redshift bins $z<0.36$, $0.36<z<0.59$ and $z>0.59$, each holding 41-42 galaxies. These sub-samples are shown in the other panels of Fig.~\ref{resres} in comparison to the local fit and scatter. For the distant data, TFR offsets $\Delta M_B$ and VSR offsets $\Delta \log \rd$ were computed as described in Sects.~\ref{anatfr} and \ref{anavsr}. To interpret these graphs, it has to be kept in mind that any evolution in luminosity and/or size is imprinted on the correlation between $\Delta M_B$ and $\Delta \log \rd$ explained above.

We find that, towards higher redshifts, the distant galaxies are gradually shifting away from the local $\Delta M_B$--$\Delta \log \rd$ relation. Using fixed-slope fits to determine the offsets from the local relation in terms of $\Delta \log \rd$, we infer an evolution of $-0.08\pm$0.06\,dex, $-0.19\pm0.08$ dex and $-0.29\pm0.07$\,dex at redshifts $z<0.36$, $0.36<z<0.59$ and $z>0.59$, resp. These offsets are depicted by solid lines in each of the distant data panels of Fig.~\ref{resres}. These deviations from the local relation are larger than the evolution in size \emph{alone} (see Sect.~\ref{anavsr}), since here $\Delta \log \rd$ is a combination of the evolution in size \emph{and} (in projection) luminosity. The scatter in the distant $\Delta M_B$--$\Delta \log \rd$ relation is larger than locally: our data yield 0.22\,dex at $z<0.36$, 0.18\,dex at $0.36<z<0.59$ and 0.20\,dex at $z>0.59$. Similar to the situation in TFR and VSR space, the scatter in the distant data hence is approx.~doubled compared to the local reference which shows $\sigma = 0.09$\,dex.

The fact that the correlation between $\Delta M_B$ and $\Delta \log \rd$ holds up to redshifts $z \approx 1$ has consequences for the combined evolution in luminosity and size of \emph{individual} galaxies. This becomes particularly clear in the $z>0.59$ bin, which represents the longest lookback times ($5.6\,{\rm Gyr}<t_{\rm lookback}<7.7\,{\rm Gyr}$) and is most sensitive to the evolution with cosmic time. The shape of the distribution does not appear to be changed with respect to the local universe, but merely shifted towards higher luminosities and smaller sizes. The galaxies which evolved strongest in luminosity are \emph{not} the ones that evolved strongest in size, and vice versa. The galaxies with the strongest decrease in size scatter around $\Delta M_B\approx0$, and the galaxies with the strongest evolution in luminosity mostly show $\Delta \log \rd > 0$ and hence are larger than their local counterparts at the same $\vm$. 

We finally want to address the scatter of the intermediate-redshift scaling relations. The observed scatter at $z \approx 0.5$ ($\sigma_{\rm obs} = 1.28$\,mag and $\sigma_{\rm obs} = 0.27$\,dex for the TFR and VSR, respectively)
is approx.~doubled with respect to the local reference data.
The TFR scatter we find is smaller than in the sample of
Weiner et al.~(\cite{wein06}, $\sigma_{\rm obs} \approx 1.5$\,mag),
similar to Fern\`andez Lorenzo et al.~(\cite{fern10}, 
$\sigma_{\rm obs} \approx 1.2$\,mag), and slightly larger than
the value given by Bamford et al.~(\cite{bamf06}, 
$\sigma_{\rm obs} \approx 1.0$\,mag);
all studies at redshifts similar to our data.
Part of the observed distant scatter stems from the uncertainties of the
$\vm$ derivation and the observational limitations, such as  beam smearing, 
limited spatial resolution and so forth.
We now want to clarify whether the increased scatter is driven by the
measurement errors of the galaxy parameters, or due to an increase of the
\emph{intrinsic} scatter. The observed scatter $\sigma_{\rm obs}$ of the TFR
comprises contributions from the errors $\sigma_{\rm mb}$ on the absolute 
magnitudes, errors  $\sigma_{\rm vmax}$ on the maximum rotation velocities and
the intrinsic scatter $\sigma_{\rm int}$ such that:
\begin{equation}
\sigma^2_{\rm obs} = \sigma^2_{\rm mb} + c^2 \sigma^2_{\rm vmax} + 
\sigma^2_{\rm int},  
\end{equation}
where $c$ is the TFR slope.
Our kinematic sample of 124 galaxies  
yields an intrinsic scatter $\sigma_{\rm int} \approx 1.1$\,mag. 
In comparison to other studies at similar redshifts,
our result is only slightly larger than the value given by 
Bamford et al.~(\cite{bamf06},  $\sigma_{\rm int} \approx 0.9$\,mag), 
but much larger than the $\sigma_{\rm int} \approx 0.7$\,mag 
derived by Miller et al.~(\cite{mill11}). 
Tully et al.~(\cite{tull98}) do not give the intrinsic scatter for their
local TFR sample, but the observed $B$-band scatter $\sigma_{\rm obs}=0.55$\,mag
suggests that the intrinsic scatter is of the order 
$\sigma_{\rm int} \approx 0.3$-$0.4$\,mag.
Our analysis as well as those of Bamford et al.~and Miller et~al.~hence 
show an increased intrinsic TFR scatter at intermediate redshifts.
This increase might be due to, e.g., a larger contribution of 
non-circular motions, i.e.~kinematically ``heated" disks 
(e.g., F\"orster-Schreiber et al.~\cite{foer09})  
or more frequent mismatches between photometric and kinematic position angle 
(e.g.~Kutdemir et al.~\cite{kutd10}) than locally.

In principle, it would also be possible that the stellar populations have an effect on the intrinsic TFR scatter, e.g.~due to a broader distribution in stellar $M/L$ ratios towards higher redshifts. To investigate this, we computed the stellar $B$-band $M/L$ ratios from the absorption-corrected rest-frame $(B-R)$ colors following Bell \& de Jong~(\cite{bell01}) for all galaxies in our sample. We define three redshift bins, using only the 47 galaxies with stellar masses $M_\ast > 10^{10} M_\odot$ (which we detect up to redshift $z \approx 1$), to minimize the impact of the correlation between stellar $M/L$ ratio and galaxy stellar mass.
The three resulting redshift bins with 15-16 galaxies each have median redshifts of 
$\langle z \rangle = 0.36$, $\langle z \rangle = 0.63$ and 
$\langle z \rangle = 0.84$. 
We find an r.m.s. of the $B$-band stellar mass--to--light ratio of
$\sigma_{M/L}=0.98\, [(M/L_B)_\odot]$ in the lowest redshift bin, 
$\sigma_{M/L}=0.46\, [(M/L_B)_\odot]$ at $\langle z \rangle = 0.63$ and
$\sigma_{M/L}=0.88\, [(M/L_B)_\odot]$ at $\langle z \rangle = 0.84$.
At least as far as stellar mass--to--light ratios are concerned, we thus
find no indication
that the evolution in the intrinsic TFR scatter might be (partly) driven
by the stellar populations.

\section{Conclusions\label{concl}}

Utilizing the FORS instruments of the ESO Very Large Telescope, we have 
constructed a sample of 124 disk galaxies up to redshift $z \approx 1$
with determined maximum rotation velocity $\vm$. Structural parameters such as
disk inclination, scale length etc.~were derived on HST/ACS images.
We analyzed the distant $\vm$ - luminosity (Tully-Fisher) and
$\vm$ - size relations and compared them to reference samples in the local
universe. Our main findings can be summarized as follows:
\begin{enumerate}
\item At given $\vm$, disk galaxies are more luminous (in rest-frame $B$-band)
and smaller (in rest-frame $I$-band)
towards higher redshifts. By $z=1$, we find a brightening of 
$\Delta M_B \approx -1.2$\,mag in absolute $B$-band magnitude 
and a decrease in size by a factor of $\sim$\,1.5.
\item The scatter in the Tully-Fisher and 
velocity-size relations at $z\approx0.5$ 
is increased by a factor of approx.~two with respect to
the local universe. 
\item The observed evolution in luminosity and size over 
the past $\sim$\,8\,Gyr is in good agreement with predictions from numerical 
simulations (e.g., Portinari \& Sommer-Larsen~\cite{port07},
Dutton et al.~\cite{dutt11}). 
\item An analysis of the combined evolution in luminosity and size reveals that the galaxies which show the strongest evolution towards smaller sizes at $z \approx 1$ are not those which feature the strongest evolution in luminosity, and vice versa. 
The galaxies with the strongest deviations from the local VSR towards smaller disks have luminosities compatible with the local TFR, while the galaxies with the strongest evolution in luminosity are slightly larger than their local counterparts at similar $\vm$.
\end{enumerate}

In the next paper of this series, we will conduct a comparison between
the kinematics of distant disk galaxies observed with
2-D (slit) and 3-D (integral field unit) spectroscopy,
in particular with respect to scaling relations like the TFR and VSR
(B\"ohm et al., in prep.).
Another paper will focus on the evolution of the correlation between the
maximum rotation velocity of the disk
and the stellar velocity dispersion in the bulge (B\"ohm et al., in prep.).

\begin{acknowledgements}
We thank the anonymous referee for a detailed report that was very helpful in improving the manuscript.
AB is grateful to the Austrian Science Fund (FWF) for
funding (projects P19300-N16 and P23946-N16). The authors thank 
B.~B\"osch (Innsbruck) for providing a python code of the Kelson~(\cite{kels03})
method for improved sky subtraction in the FORS2/MXU spectra.
This publication is supported by the Austrian Science Fund (FWF).
\end{acknowledgements}

\appendix

\section{The distant Tully-Fisher relation slope\label{slope}}
We implicitly assumed in our analysis in Sect.~\ref{anatfr} that the TFR slope remains constant over the redshift range $0<z<1$. We want to justify this in the following. To derive the distant TFR slope, we rely on the so-called inverse fit which is of the form
\begin{equation}
\log \vm = f ( M_B ) = a \cdot M_B + b. \label{tfinv}
\end{equation}
This fitting method is more robust against selection effects arising
from a magnitude limit than the classical forward fit 
\begin{equation}
M_B = f(\vm) = c \cdot \log \vm + d \label{tfforw},
\end{equation}
as has been demonstrated e.g.~by Willick et al.~(\cite{will95}).
Even though some statements given in an earlier work of this author 
(Willick~\cite{will94}) imply that the inverse fitting method is prone to a 
magnitude bias of similar
strength as the forward fitting method, the results of  
Willick et al.~(\cite{will95}) clearly show that the impact of the ``magnitude
bias" arising from sample incompleteness towards lower luminosities is 
much weaker when an inverse fit is utilized. In fact, the strength of the
magnitude bias is reduced by a factor of six when using the inverse fit,
``reducing the bias from a significant concern to a marginal effect"
(Tully \& Courtois~\cite{tull12}).
 
Using the parametrization in Eq.~\ref{tfinv}, we find 
\begin{equation}
\log \vm = (-0.131 \pm 0.012 ) \cdot M_B - (0.494 \pm 0.244) 
\end{equation}
for our full data set. This corresponds to a distant TFR slope of $c=-7.62^{+0.63}_{-0.78}$ in the form of Eq.~\ref{tfforw} and agrees well with the slope of $c=-7.79$ for the local sample of Tully et al.~(\cite{tull98}), who also used the inverse method. We hence find that the slope of the $z \approx 0.5$ TFR is compatible with the local one. Most other observational studies also derived (or assumed) a TFR slope independent of look-back time (e.g., Bamford et al.~\cite{bamf06}, Miller et al.~\cite{mill11}; note that these authors also relied on an inverse fit for their analysis).
Only Weiner et al.~(\cite{wein06}) reported on an increased TFR slope in the range $0.4<z<1$; however, their results are not directly comparable to ours since the approach by Weiner et al.~lacks corrections for disk inclination and, in turn, for the inclination-dependent optical beam smearing effect (see Sect.~\ref{anakin}).

In the analysis of an earlier stage of our kinematic survey, we found a shallower TFR slope at $z \approx 0.5$ using a bisector fit (B\"ohm et al.~\cite{boeh04}) and showed that an apparent slope evolution in a magnitude-limited survey could be \emph{mimicked} by a strong increase of the TFR scatter with look-back time (B\"ohm \& Ziegler~\cite{boeh07}). Applying a bisector fit to our current sample, we find a slope of $c=-5.02 \pm 0.47$. A forward fit (Eq.~\ref{tfforw}) would yield an even shallower slope of $c=-3.71 \pm 0.35$. However, both these fitting methods are sensitive to the impact of the magnitude bias. To demonstrate this, we use the methodology introduced by Giovanelli et al.~(\cite{giov97}) to perform a correction of the magnitude bias (as carried out also in B\"ohm \& Ziegler~\cite{boeh07}). In this approach, the observed luminosity distribution is compared to a Schechter luminosity function to infer the sample completeness at a given magnitude and, in turn, a given $\vm$. The key factor governing the impact of the magnitude bias is the TFR scatter, for which we use the observed scatter of our whole sample, i.e., $\sigma_{\rm obs}=1.28$\,mag at $z\approx0.5$. After the de-biasing procedure, the absolute magnitudes are less bright; in particular, for galaxies at low $\vm$. For the corrected sample, we find slopes of $c=-5.82\pm0.35$ (forward fit), $c=-6.86\pm0.41$ (bisector fit) and $c=-8.33^{+0.47}_{-0.53}$ (inverse fit). These numbers, when compared to the fits of the uncorrected sample, clearly demonstrate that the inverse TFR fit is the least sensitive to the influence of sample incompleteness. In fact, the inverse fit slopes of corrected and uncorrected sample agree within the errors. If we would apply the de-biasing procedure also to the local sample of Tully et al. (which most likely would lead only to very small changes of the absolute magnitudes and, in turn, only a very small steepening of the local TFR slope; however, we could not do this exercise since the data are not electronically available), local and distant inverse-fit slopes would most probably still be in agreement within the fit errors.

Note that we find large differences between a forward, bisector and inverse TFR fit not only for our sample but also other studies at similar redshifts. Using the sample from Bamford et al.~(\cite{bamf06}), which comprises 89 field disk galaxies at $0.06<z<1.0$, we infer the following TFR slopes: $c=-4.59\pm0.44$ (forward), $c=-5.89\pm0.56$ (bisector) and $c=-8.18^{+0.71}_{-0.86}$ (inverse). Alternatively using the sample from Miller et al.~(\cite{mill11}), which contains 129 disk galaxies at $0.20<z<1.31$, we find these slopes: $c=-3.99\pm0.45$ (forward), $c=-5.79\pm0.65$ (bisector) and $c=-7.90^{+0.62}_{-0.74}$ (inverse). This demonstrates that strong differences between the three fitting methods are common for distant samples, motivating the use of the method which by far is least sensitive to the magnitude bias, i.e., the inverse TFR fit.

\end{document}